# A Bi-Level Cooperative Driving Strategy Allowing Lane Changes

Huile Xu [1, 2], Yi Zhang [1, 3], Christos G. Cassandras [2], Li Li [1, *], Shuo Feng [4]

1. Department of Automation, Tsinghua University, Beijing, China 100084
2. Division of Systems Engineering and Center for Information and Systems Engineering, Boston University, Boston, MA, USA 02215
3. Tsinghua-Berkeley Shenzhen Institute (TBSI), Shenzhen, China 518055
4. Department of Civil and Environmental Engineering, University of Michigan, Ann Arbor, MI, USA 48109

*Abstract*: This paper studies the cooperative driving of connected and automated vehicles (CAVs) at conflict areas (e.g., non-signalized intersections and ramping regions). Due to safety concerns, most existing studies prohibit lane change since this may cause lateral collisions when coordination is not appropriately performed. However, in many traffic scenarios (e.g., work zones), vehicles must change lanes. To solve this problem, we categorize the potential collision into two kinds and thus establish a bi-level planning problem. The right-of-way of vehicles for the critical conflict zone is considered in the upper-level, and the right-of-way of vehicles during lane changes is then resolved in the lower-level. The solutions of the upper-level problem are represented in tree space, and a near-optimal solution is searched for by combining Monte Carlo Tree Search (MCTS) with some heuristic rules within a very short planning time. The proposed strategy is suitable for not only the shortest delay objective but also other objectives (e.g., energy-saving and passenger comfort). Numerical examples show that the proposed strategy leads to good traffic performance in real-time.

1. Introduction

    Increasing traffic congestion and accidents have caused huge losses to society and generated

wide concern in recent years (Rios-Torres *et al*., 2016). The emergence of Connected and Automated Vehicles (CAVs) and CAV-based traffic control is believed to be a promising way of improving safety and traffic efficiency. With the aid of vehicle-to-everything (V2X) communication, CAVs can share their driving states (position, velocity, acceleration, etc.) and intentions with adjacent vehicles and road infrastructure to better coordinate their motions (Li *et al*., 2014; Sukuvaara *et al*., 2009).

The existing studies for CAV-based traffic control can be categorized into six types of approaches (Guo *et al*., 2019), that is, driver guidance (Ubiergo and Jin, 2016), actuated (adaptive) signal control (Yun and Park, 2012), platoon-based signal control (Lioris *et al*., 2017), planning-based signal control (Goodall *et al*., 2013), signal-vehicle coupled control (Yu *et al*., 2018), and multi-vehicle cooperative driving without traffic signals (Chen and Englund, 2016). Different from the other five types of approaches, most studies about multi-vehicle cooperative driving requires a 100% CAV environment but does not rely on the traffic signal control system like traffic lights and stop signs. In recent years, there also have been some studies to investigate the performance of cooperative driving under different penetration rates, revealing that multiple benefits still can be offered by CAVs even there are some human-driving vehicles in the environment (Zhang and Cassandras, 2019). Thus, with the rapid development of CAVs, it is regarded as the most promising and efficient intelligent transportation system (ITS) in the future.

The main task of cooperative driving is to cooperatively control CAVs passing through the conflict areas safely and efficiently without any traffic signaling. The concept of cooperative driving first appeared in the early 1990s. The Association of Electronic Technology for Automobile Traffic and Driving presented it for flexible platooning of automated vehicles with a short inter-vehicle distance (Tsugawa, 2002). Since then, cooperative driving has been continuously studied by many researchers and examined by various projects, e.g., the Demo 2000 Cooperative Driving System in Japan (Kato *et al.*, 2002) and the Grand Cooperative Driving Challenge in Netherlands (Englund *et al.*, 2016). Researchers found that one of the key points for cooperative driving was to determine the right-of-way of vehicles for conflicting areas (Li *et al.,* 2006; Guler *et al.,* 2014). An assignment of the right-of-way of vehicles for the critical conflict zone generates a possible passing order for vehicles.

As summarized in (Meng *et al.*, 2017), there are two primary kinds of cooperative driving strategies, ad hoc negotiation-based and planning-based, for determining the passing order.

Ad hoc negotiation-based strategies aim to assign right-of-way using some heuristic rules within a very short time (Xu *et al.*, 2019a). Dresner and Stone (2008) proposed a reservation-based intersection management strategy which divided the intersection into grids (resources) and assigned these grids to CAVs in a First-In-First-Out (FIFO) manner. Choi *et al.* (2018) extended the idea of reservation-based cooperative traffic management to an intersection of multi-lane roads. They considered moving directions of vehicles when passing through the intersection and found that the vehicles turning left greatly contribute to the overall average delay. Malikopoulos et al. (2018) proposed a decentralized energy-optimal control framework for CAVs at signal-free intersections in which the right-of-way (desired arrival times to the intersection) was determined according to the FIFO manner and the trajectories (velocity and acceleration profiles) of vehicles were derived through a decentralized optimal control problem. To realize a fast implementation, they presented a complete analytical solution for the decentralized optimal control problem. For simplicity, the considered scenario in the work was an isolated single-lane intersection with no lane changes and no turns allowed. Then, Zhang et al. (2018) further extended the decentralized energy-optimal control framework by including left and right turns and proposed a dynamic resequencing method for relaxing the FIFO constraints and exploring some other possible right-of-way. Besides the intersection scenarios, the FIFO-based passing order can be easily extended to resolve the conflicts in other traffic scenarios such as highway ramps and work zones. Other conflict resolutions for assigning the right-of-way like conflict graph (Liu et al., 2018) and virtual vehicles (Uno et al., 1999; Xu et al., 2018) also have been attempted to be applied in this field. However, as shown in (Meng et al., 2017), the passing order found by ad hoc negotiation-based strategies roughly followed the FIFO rule and were not good enough in many situations.

Planning-based strategies aim to enumerate all possible passing orders to find a globally optimal solution (Xu et al., 2019a). Most state-of-the-art studies formulate the problem as an optimization problem whose objective is usually set to minimize the total delay or passing time of all CAVs (Meng et al., 2017; Li et al., 2017). Li et al. (2017) formulated the intersection automation policy within a 100% CAV environment as a mixed-integer linear programming (MILP) problem whose decision variables were desired arrival times and used the branch-and-bound search approach to find the exact optimal solution. Hult et al. (2016) formulated an optimal coordination problem for vehicles and the decision variables were states and control signals for each vehicle. They used model predictive control to solve the problem and took a simple intersection scenario with six vehicles as an example. Apart from the above, the objective of some studies is to minimize the overlap of vehicle trajectories inside the intersection zone. For example, Kamal et al. (2014) proposed a vehicle-intersection coordination scheme for preventing each pair of conflicting vehicles from approaching their cross-collision point at the same time. A risk function was designed to indicate the risk of a collision of a pair of vehicles, and then the model predictive control was used to solve the resulting constrained nonlinear optimization problem. The tree search method is an equivalent formulation to the optimization method. Li et al. (2006) showed that we can also view the cooperative driving problem as a tree search problem. Each tree node indicates a special passing order and the equivalent objective was to find the node corresponds to the minimum objective value. However, all these studies ignore the lane change for the sake of safety concerns and simplicity. Moreover, the computation time of all planning-based strategies increases sharply as the number of vehicles increases (Lawler et al., 1966; Morrison *et al.*, 2016). This hinders their applications in practice.

In recent years, some state-of-the-art studies have started taking lane changing into consideration mainly because a) no feasible solution exists for collision avoidance in some driving scenarios without considering a lane change; and b) the latest development of CAV technology is beginning to meet the requirement of control and positioning accuracy for lane changes. Lu *et al.* (2019) considered lane changing and formulated the traffic management of vehicle trajectories as a mixed-integer nonlinear programming (MINLP) problem and developed a specialized algorithm based on the rolling horizon approach to improve the computational efficiency. They aimed to optimize both longitudinal and lateral trajectories for all vehicles, subject to vehicle kinematics and collision avoidance. For simplicity, they assumed that a lane change maneuver can always be completed within a given time interval. Hu *et al.* (2019) showed that when the passing order of vehicles in the cooperative lane change region was determined, the trajectories of vehicles could be efficiently optimized by a linear programming model. The FIFO-based passing order was used, and the lane change maneuvers of all vehicles were assumed to be completed within the same time interval. Nevertheless, the assumption is practically impossible, and the lane change trajectory used is not mentioned. Smooth lane change trajectories approximated by the sixth order polynomials were considered in (Li *et al.*, 2005). They showed that the lane change trajectories of vehicles can be considered and optimized through a constructed tree search problem. Then, this idea was further extended from intersections to lane closures (Li *et al.*, 2007). Thus, the lane change can be carried out in a more practical way. However, it is difficult to find a good solution when the number of vehicles is large. Although the lane change was not considered, Xu *et al.* (2019b) showed that Monte Carlo tree search with heuristic rules can help us to search a good solution even when the search space is huge. As seen from the above, the studies of cooperative driving allowing lane changing is limited or oversimplified.

In summary, there are two problems to conquer in this research direction. First, it is difficult to handle lane changing in local conflict areas and deal with the high nonlinearity caused by considering the lane change trajectories. Second, the following vehicle may pass through the conflict zones earlier than the preceding vehicle because of the lane change, which results in the sharp increase of the size of the search space for possible passing orders. The increasing number of the passing orders makes it difficult to find a good enough passing order within limited computation time. To realize a fast implementation, many studies use FIFO-based rule or other heuristic rules to assign right-of-way to vehicles. However, the performance of the solution cannot be guaranteed.

To address these two major limitations, we propose a bi-level-based cooperative driving strategy allowing lane changes. According to the two types of potential collisions, we establish a bi-level planning problem in which the optimization problem for the cooperative driving is broken down into two sub-problems. For the first problem of finding the optimal passing order, we creatively build a tree representation of the solution space for passing orders. After that, we combine the Monte Carlo Tree Search (MCTS) with some heuristic rules to find a more promising passing order than the FIFO-based passing order. For the second problem of deriving the objective value and the corresponding trajectories of vehicles, we design a passing-order-to-trajectory interpretation algorithm to quickly derive a feasible solution for the optimization problem on condition that the passing order is given. For each lane change vehicle, a suitable lane change trajectory constrained by vehicle dynamics is chosen from a pre-designed trajectory set according to the velocity of the vehicle. Thus, the right-of-way of vehicles for the critical conflict zone is considered in the upper-level, and the right-of-way of vehicles during lane changes is then resolved in the lower-level. Testing results show that the proposed strategy can effectively improve traffic efficiency with a short enough computation time.

The main contributions of the paper include: (a) we explicitly consider the lane change trajectories constrained by vehicle dynamics and make the solution framework more practical than most of the existing studies; (b) we formulate a bi-level planning which greatly reduces the complexity of the problem and makes it more efficiently solved than the conventional MIP problem; (c) we use the MCTS-based tree search to realize the trade-off between the computational efficiency and coordination performance.

To give a better presentation of our finding, the rest of this paper is arranged as follows. *Section 2* introduces the problem and formulates it into an optimization problem. *Section 3* briefly reviews the existing strategies and presents our new strategy. *Section 4* shows the testing results of this new strategy. Finally, *Section 5* gives conclusions.

## 2. Problem Presentation

### 2.1 Conflict Zone Classification

We can classify all conflict zones into two kinds, critical conflict zones and local conflict zones, according to the impact of the conflict on the driving scenario.

Critical conflict zones are the relatively fixed areas where a lot of potential collisions may occur if vehicles do not appropriately adjust their motion. They are usually the bottleneck area of traffic flow and are caused by two main reasons: 1) changes in road geometry and 2) unreasonable occupancy of road resources. For the former, lane crossing and lane merging are the most common causes of road geometry changes. The resulting increase in collision risk and the reduction of the maximum traffic capacity require good coordination of the motions of the vehicles. For the latter, the unreasonable occupancy of road resources often leads to the reduction of the number of lanes and blocks the traffic flow in several lanes. For example, a part of the road in a lane is disabled because of a work zone, an accident vehicle, or other obstacles. All vehicles in the same lane where a part of the road is disabled must change their lanes in advance to avoid a collision with them. The classic critical conflict zones include intersections, highways with a work zone, etc.

In contrast, local conflict zones are the areas where conflicts are caused by the maneuvers of vehicles. These local conflicts influence the motions of several adjacent vehicles for a small period of time. For instance, the local conflict zone may appear due to a lane change maneuver of one vehicle and quickly disappear when the maneuver is completed.

Most of the complicated driving scenarios are made up of several basic driving scenarios, like intersection, on-ramp, off-ramp, roundabout, road segment, etc. For example, a simple urban traffic network can be divided into several intersection scenarios and road segment scenarios based on the geometry of the road. As shown in Fig. 1, most basic driving scenarios include only one critical conflict zone which lies in the downstream of local conflict zones. Thus, if we can handle the conflicts in all basic driving scenarios well, then it can be extended to deal with all complicated driving scenarios like traffic networks by dividing the conflicts into each basic driving scenarios. The main goal of the paper is to propose an effective way to solve the conflicts in basic driving scenarios.

Since the critical conflict zones have significant impact on the traffic efficiency of the driving scenario, the primary task for cooperative driving is to optimize the assignment of right-of-way in the critical conflict zone. We can first plan the motions of vehicles that are approaching to the critical conflict zone and then backwardly plan their motions in the local conflict zone.

Yang *et al.* (2016) pointed out that all kinds of conflicts arise from the unclear assignment of the right-of-way. For the critical conflict zone, we define a passing order to be a possible result of assigning right-of-way and represents the priority of vehicles. For simplicity, we can use a string to denote a passing order (Li *et al.*, 2006). For example, string BAC means vehicle B, vehicle A, and vehicle C enter into the critical conflict zone sequentially. The vehicle with the leftmost position in the string has the highest priority. If there is a potential collision between two vehicles, no matter in either the critical or local conflict zone, the vehicle with higher priority in the passing order can pass through the conflict zone first.

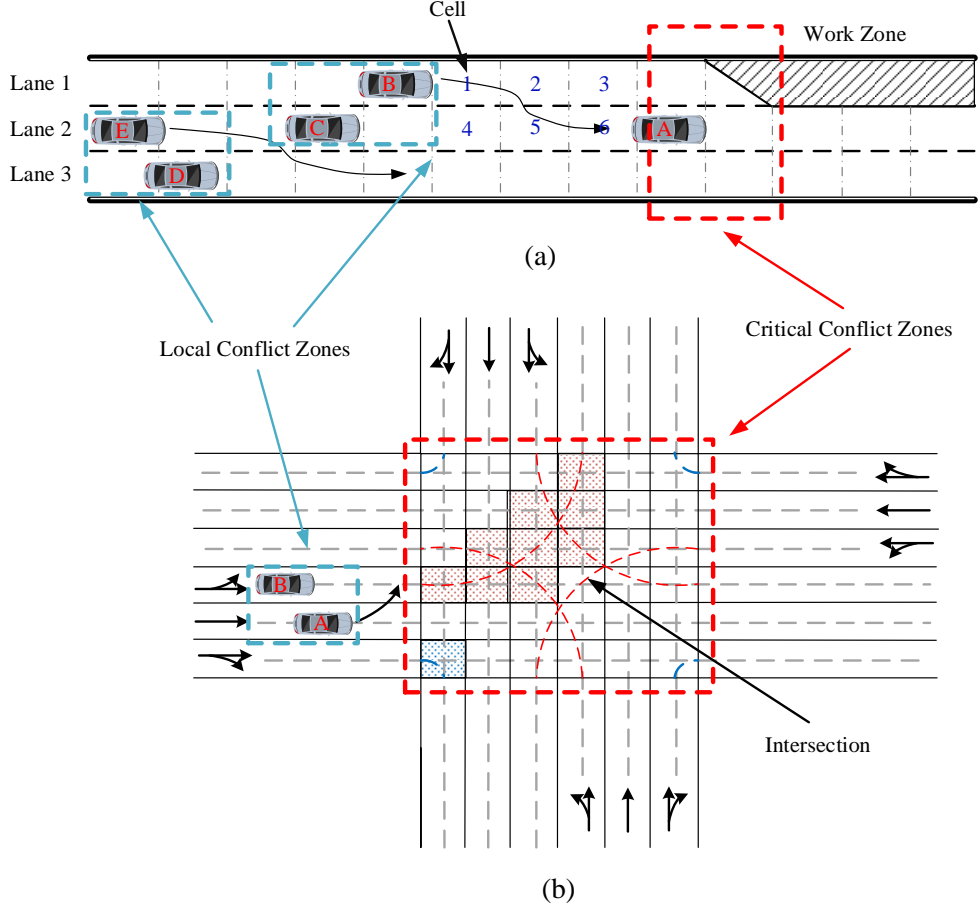

Fig. 1. Typical driving scenarios with conflict zones. The red dashed boxes represent critical conflict zones, and the blue dashed boxes represent local conflict zones. (a) Vehicle B changes lanes to avoid a collision with the work zone; (b) The left-turning vehicle A changes its lane to the left lane because of driving rules.

## 2.2 The Trajectory Planning Problem

If we take the locations of vehicles as decision variables, we can formulate the cooperative driving problem as a trajectory planning problem. The constraints include vehicle dynamics, collision avoidance, and physical constraints. The trajectory planning problem can be formulated as

$$\min_{x(t),u(t)} \sum_{i=1}^{n} J_i(x_i(t), u_i(t)) \tag{1a}$$

$$s.t. \quad \dot{x}_i(t) = f_i(x_i(t), u_i(t), t) \tag{1b}$$

$$g(x_i(t), x_j(t), b_{i,j}) = 0 \tag{1c}$$

$$v_{i,\min} \leq v_i(t) \leq v_{i,\max} \tag{1d}$$

$$u_{i,\min} \leq u_i(t) \leq u_{i,\max} \tag{1e}$$

$$b_{i,j} \in \{0,1\} \tag{1f}$$

where $x(t)$ and $u(t)$ represent the state and control input for each vehicle at time $t$. The state $x(t)$ is a vector that consists of the vehicle's position and velocity. The function $J_i(\cdot)$ is an objective function which could be very general including traffic efficiency, energy consumption, etc. Since we are more concerned about the traffic efficiency, we limit ourselves to an objective which is the delay. Therefore, the objective function in (1a) is the total delay. Of course, by dividing by the number of vehicles n, we get the average vehicle delay. It is worth noting that the number of vehicles in the control zone varies with time, and we solve the above trajectory planning problem in a time-driven manner like every few seconds. Every time we formulate the problem, n is the number of vehicles in the control zone at that time. The function $f_i(\cdot)$ in (1b) is the vehicle dynamics constraint for vehicle $i$, the function $g_i(\cdot)$ in (1c) is the collision avoidance constraints, and (1d) and (1e) is the physical constraints for velocity and control input. The binary variable $b_{i,j}$ is introduced to denote the right-of-way priority between vehicle $i$ and vehicle $j$. For example, if $b_{i,j}$ equals 0, the vehicle $i$ has higher right-of-way than vehicle $j$; otherwise, the vehicle $j$ has higher right-of-way.

The optimization problem (1) is a continuous-time optimal control problem. For practical implementation, we can discretize the control horizon and use the locations of vehicles at each sampling time as decision variables, which results in a similar discrete-time optimal control problem. When the sampling time is small enough, the resulting trajectories are still smooth and continuous (Li and Li, 2019).

Clearly, the formulated optimization problem is a MINLP that is hard to solve. Although there are many algorithms for collision avoidance constraints (1c) (Mukhtar *et al.*, 2015; Ji *et al.*, 2017), it is still time-consuming to do collision avoidance for vehicles especially when the lane change is considered.

To solve the problem, some studies use the cell-based modeling method and discretize the road space into several cells as shown in Fig. 1 (Dresner and Stone, 2004; Dresner and Stone, 2008). The length of the cells can be determined according to the length of the vehicles. For simplicity, we assume that all vehicles are homogeneous passenger cars, and the lengths of all cells are the same. The lane change vehicles are assumed to simultaneously occupy the cells in both the origin and destination lane of the maneuver. Then the safety in the critical conflict zone and local critical zones can both be guaranteed by requiring that every cell is occupied by no more than one vehicle at any time. It can be easily realized through optimizing the arrival times to all cells for all vehicles.

The decision variables in this method are the arrival times of vehicles instead of the locations of vehicles. Compared with the original formulation, the collision avoidance constraints of the cooperative driving problem can be further discretized and simplified as

$$t_i(z) - t_j(z) + M \cdot b_{i,j} \geq \Delta t \quad \forall i, j, z,  \tag{2a}$$

$$t_j(t) - t_i(t) + M \cdot (1 - b_{i,j}) \geq \Delta t \quad \forall i, j, z,  \tag{2b}$$

where $t_i(z)$ is the arrival time of the vehicle $i$ to the $z$ th cell, $M$ is a sufficiently large number and $\Delta t$ is the safety headway. Clearly, when $b_{i,j}$ equals 0, the constraint (2a) is reduced to

$$t_i(z) - t_j(z) \geq \Delta t \quad \forall i, j, z, \tag{3}$$

and the constraint (2b) is not active because of the large $M$. So, vehicle $i$ must arrive at the $z$ th cell later than vehicle $j$, that is to say, vehicle $j$ has higher right-of-way than vehicle $i$. Otherwise, when $b_{i,j}$ equals 1, vehicle $i$ has higher right-of-way. Thus, the vector consisting of all binary variables $\boldsymbol{b} = [b_{1,2}, b_{1,3}, \cdots b_{i,j}, \cdots b_{n-1,n}]$ is an equivalent expression to the passing order and represents the right-of-way priority for all vehicles.

If the temporal and spatial duration of the lane change trajectory is not considered, the constraint (1b) under cell discretization can be written as

$$t_i(z+1) = f_i^1(t_i(z), u_i(z), v_i(z), z), \tag{4}$$

where $f_i^1(\cdot)$ is the function that describes the time constraint between any two adjacent cells. For example, if we only consider the longitudinal movement of a vehicle and assume that the acceleration of a vehicle in a cell remains the same, we have

$$\begin{aligned} v(t_i(z))(t_i(z+1) - t_i(z)) + 0.5u(t_i(z))(t_i(z+1) - t_i(z))^2 = \Delta p \\ \Downarrow \\ t_i(z+1) = \frac{\sqrt{v(t_i(z))^2 + 2u(t_i(z))\Delta p} - v(t_i(z))}{u(t_i(z))} \end{aligned} \tag{5}$$

where $\Delta p$ is the length of a cell.

However, to be more realistic, we consider the lane change trajectory instead of assuming the lane change can be completed immediately or within the distance of one cell. The lane change trajectory satisfies vehicle dynamics and can be designed in advance using fifth-order polynomial curve functions or other functions that can guarantee a continuous third derivative and smooth curvature (Li *et al.*, 2005; Papadimitriou and Tomizuka, 2003; Wang *et al.*, 2014; Luo *et al.*, 2016).

For vehicles during the lane change process, the constraints (4) should be modified since the lane change trajectory is determined in advance and the input $u(t)$ is given correspondingly. For example, as shown in Fig. 1, the lane change maneuver of vehicle B begins at the start-edge of cell 1 and ends at the end-edge of cell 6. Based on the given lane change trajectory, suppose that the vehicle B spends 2s, 1s, and 1.5s passing through cell 1, cell 2, and cell 3 respectively. Then, we have

$$t_B(2) = t_B(1) + 2, \tag{6}$$

$$t_B(3) = t_B(2) + 1. \tag{7}$$

As aforementioned, lane change vehicles simultaneously occupy the cells in two lanes, that is,

$$t_B(4) = t_B(1), t_B(5) = t_B(2), t_B(6) = t_B(3). \tag{8}$$

Thus, during the lane change process, the arrival times to cells can be directly determined according to the given lane change trajectory $\Gamma_i$, which can be expressed as

$$t_i(z+1) = f_i^2(t_i(z), \Gamma_i), \tag{9}$$

where $f_i^2(\cdot)$ is the function that indicates the arrival time gap between two adjacent cells on condition that the lane change trajectory $\Gamma_i$ is given.

Thus, the trajectory planning problem under cell discretization can be expressed as

$$\min_{t(z),u(z)} \sum_{i=1}^{n} J_i(t_i(z), u_i(z)) \tag{10a}$$

$$s.t. \quad (2a), (2b), (4), (9) \tag{10b}$$

$$v_{i,\min} \leq v_i(z) \leq v_{i,\max} \tag{10c}$$

$$u_{i,\min} \leq u_i(z) \leq u_{i,\max} \tag{10d}$$

$$b_{i,j} \in \{0,1\} \tag{10e}$$

After the lane change trajectory is considered, more binary variables need to be introduced to indicate the states (changing lane or going straight) of vehicles in cell $i$. Meanwhile, the possible trajectories of vehicles become more complicated, which makes the problem (10) difficult to be handled.

## 3. The Bi-level Planning Framework

To solve above MIP problems, researchers have proposed various methods and most of them are planning-based methods. The planning-based methods can be regarded as single level planning methods that aim at solving the formulated MIP model directly. However, the computation time is too large since the collision avoidance is complicated. Some ad hoc negotiation-based methods proposed that we can determine the passing order of vehicles first according to some heuristic rules, and then schedule the trajectories of all vehicles according to the given passing order. However, the performance of the given FIFO-based passing order cannot be guaranteed. Thus, it is necessary to propose a new strategy that can deal with different kinds of conflicts and at the same time has the ability of balancing the coordination performance and computation time.

Based on the above observation of conflict zone classification, it is natural for us to put forward a bi-level planning framework that maximizes a specific objective such as traffic efficiency in the upper-level problem through optimizing the passing order of vehicles and resolves all local conflicts in the lower-level problem, see Fig. 2.

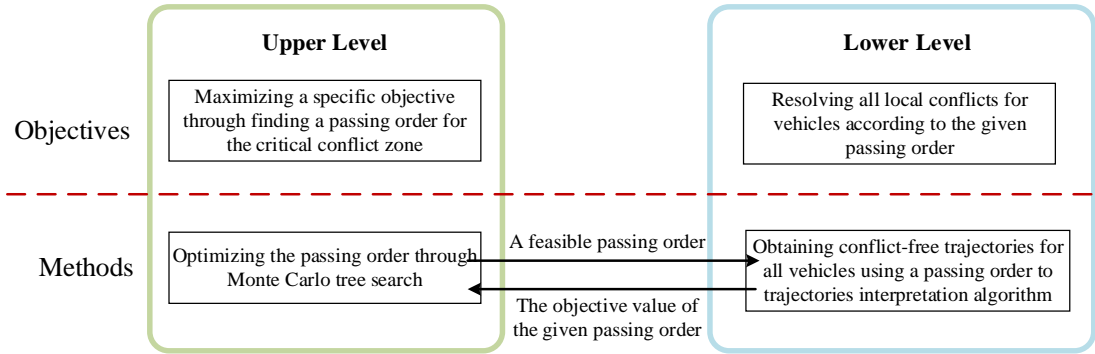

Fig. 2. Bi-level planning framework.

In the upper-level problem, we transfer the problem of finding the optimal passing order into a tree search problem (Li *et al*., 2006; Xu *et al*., 2019b). Next, we use Monte Carlo tree search with some heuristic rules to accelerate the searching process. Instead of exploring all possible passing orders like the planning-based methods, the Monte Carlo tree search method tries to search the passing orders that are promising to be the optimal passing order, and the amount of this kind of passing order only accounts for a small part of all passing orders.

In the lower-level problem, we design a passing-order-to-trajectory interpretation algorithm to quickly derive conflict-free trajectories for all vehicles according to the passing priority represented by the passing order (Malikopoulos *et al*., 2018; Xu *et al*., 2019b). The trajectory of the vehicle that has higher priority will be scheduled earlier. This algorithm provides a performance evaluation of the given passing order for the upper-level problem. It is worth noting that different passing-order-to-trajectory interpretation algorithms can lead to different performance evaluations. Our goal is to propose an algorithm that can quickly obtain an accurate solution for the lower-level problem. However, sometimes when the lower-level problem is complicated, we tend to obtain an approximate solution of the lower-level problem with a very small computation time instead of deriving an accurate solution with a large amount of computation time since the sub-optimality does not noticeably degenerate the performance of the whole bi-level planning framework.

Through solving these two problems iteratively, the current best passing order is continuously updated. When the computation budget is reached, the search process terminates and returns the best passing order and corresponding conflict-free trajectories of vehicles.

**3.1 Upper-level Planning**

As pointed out in (Meng *et al.*, 2017), we can formulate the whole problem as a tree search problem in the solution space that consists of all possible passing orders. Each leaf node represents a possible passing order of vehicles which can also be denoted as a string (Li *et al.*, 2006). Passing order indicates the priority of all vehicles and the vehicle which ranks higher in the passing order has higher priority of occupying a road resource. For example, string CAB means vehicle C has the highest priority and other vehicles need to decelerate and let it go first when there is a conflict between them. If there is no conflict between vehicle A and vehicle C, the performance of two passing orders CAB and ACB would be the same.

The following part takes the scenario shown in Fig. 1(a) as an example to show how to construct a search tree. To discard some possible passing orders that are not promising to be the optimal solution, we make the following assumptions: 1) The vehicles in the middle lane can choose going straight or changing lanes, and we use the index symbol with subscript "change lane" to represent the vehicles which will change lanes; 2) The vehicles in lane 1 must change lanes to avoid a collision with the work zone, and the vehicles in lane 3 must not change lanes to make the middle lane have more space to accommodate the vehicles from lane 1, so we do not use subscripts to indicate their actions; 3) For vehicles in the same lane, the preceding vehicle should have higher priority than the following vehicles.

Let us take the scenario shown in Fig. 1(a) as an example to explain how to interpret the tree representation of the solution space. At first, we set the passing order in the root node to be empty. Then, each direct child node of the root node (in the second layer) refers to one index symbol that indicates the first vehicle in a special passing order. Due to the above assumption 3, the first vehicle only could be the leading vehicle in three lanes shown in Fig. 1(a), i.e., vehicle A, vehicle B, and vehicle D. Moreover, a subscript "change lane" should be added to vehicle A since it may change lanes. Next, the nodes in the third layer refer to one string consisting of two indices symbols that indicate the first two vehicles in a special passing order. As shown in Fig. 3, if we choose vehicle A with going straight action as the first vehicle in the passing order, then the vehicle C becomes the new leading vehicle in lane 2, and the second vehicle in the passing order could be the vehicle B, vehicle C, and vehicle D. Similarly, the child nodes expand their child nodes, and all possible passing orders are generated as leaf nodes in the bottom layer of the solution tree as shown in Fig. 3.

It is usually impossible to expand all the nodes of the solution tree within the limited computation budget when there are lots of unplanned vehicles. Thus, we use MCTS to search nodes with the potential to be the optimal solution. The successful application of MCTS in the game of Go shows it is an effective way to deal with such problems (Silver *et al.*, 2017).

MCTS gradually builds a search tree in an iteration way, and one iteration consists of four steps including selection, expansion, simulation, and backpropagation (Kocsis *et al.*, 2006; Browne *et al.*, 2012). The detailed iteration operation is shown in Fig. 4. Due to the space limit, the detail introduction of the four steps is neglected. Interested readers can refer to (Browne *et al.*, 2012; Xu *et al.*, 2019b).

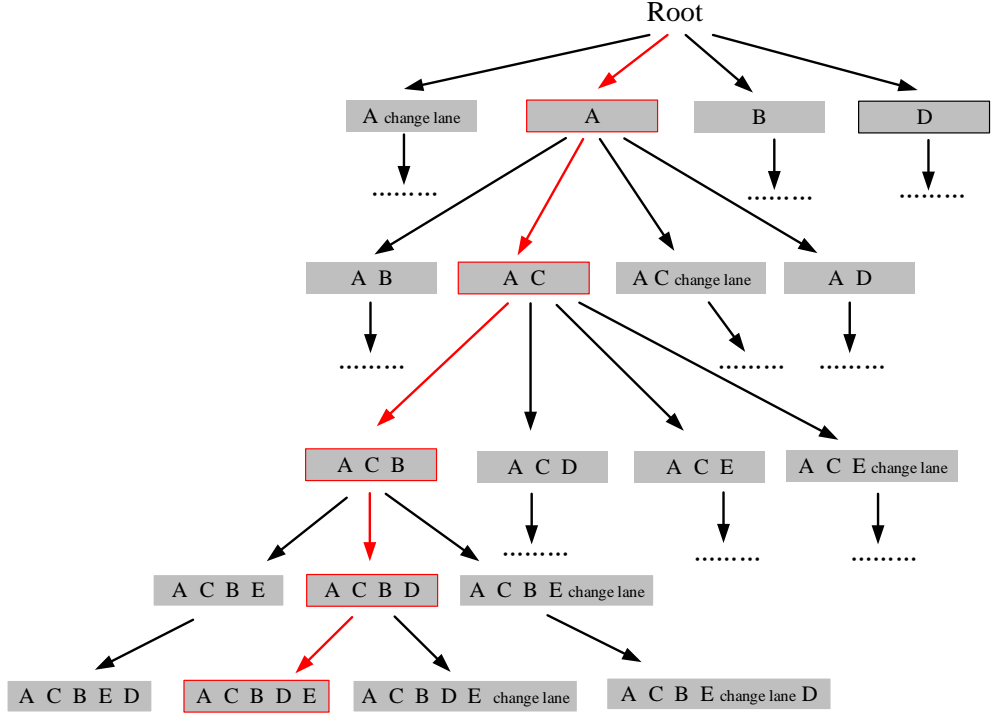

Fig. 3. The solution tree stemmed from the intersection scenario shown in Fig. 1(a). The leaf nodes in the bottom layer represent the complete passing orders for all vehicles. For vehicles in the middle lane, we use subscript "change lane" to denote whether the vehicle will change lane.

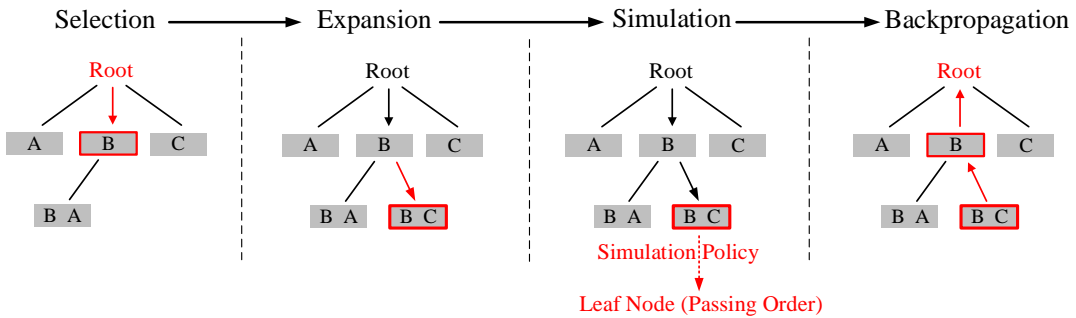

Fig. 4. One iteration of the MCTS.

During the building process of the search tree, the best passing order is continuously updated. As soon as the computation budget is reached, the search terminates and returns the state-of-the-art best passing order. The velocity and acceleration profiles of all vehicles can be calculated according to the best passing order by using a passing-order-to-trajectory interpretation algorithm that will be introduced in the lower-level planning.

The classical MCTS uses random sampling and adds the uncovered vehicles into the passing order string one by one until we find a complete passing order string and reach the maximum depth of the tree from the current new node without branching (Browne *et al.*, 2012). For example, when we apply a random sampling policy to the node BC shown in Fig. 4, we can randomly expand a direct child node in its next layer; say node BCA. The node BCA will be further expanded by repeating such a process until a leaf node (e.g., node BCADE) is expanded. Finally, the potential of the new node will be evaluated by the partial passing order and its simulated off-spring leaf nodes (passing orders).

However, the number of possible passing orders is huge especially when the lane change is considered. The passing orders generated by random sampling cannot quickly capture the real potential of a node during simulation. To solve the problem, some heuristic rules based on human knowledge are added into the simulation policy to help us to decide which nodes (vehicles) should be expanded (added into the candidate passing order string) first as shown in **Algorithm 1**.

Before we introduce the **Algorithm 1**, we first define a set $\Omega$ to include identities of all vehicles sorted by their longitudinal location. For example, $\Omega(1)$ stores the identity of the vehicle whose location is the farthest from the entrance in a longitudinal sense.

---

**Algorithm 1** Heuristic Simulation Policy

Input: The set $\Omega$

Output: A possible passing order

1) Set $i = 1$. Among all vehicles, we consider the vehicle $\Omega(i)$ first.
2) If the action of the considered vehicle is going straight, we add it into the initial passing order.
3) If the action of the considered vehicle is changing lanes, we judge whether the collision avoidance conditions are satisfied. If the result is true, we add it into the initial passing order. Otherwise, $i = i + 1$ and consider the vehicle $\Omega(i)$. Repeat the steps 2 and 3 until a vehicle is added.
4) We delete the vehicle $\Omega(i)$ from the set $\Omega$. Repeat the steps 1, 2, and 3 until a complete passing order is generated.
5) The objective value (10a) of the generated passing order can be easily derived by a passing-order-to-trajectory interpretation algorithm.

---

**3.2 Lower-level Planning**

As aforementioned, the decision variables of the optimization problem (10) are the passing order, desired arrival times, and acceleration profiles of all vehicles. If a passing order is given, we can design a passing-order-to-trajectory interpretation algorithm for directly deriving the remaining variables in the optimization problem, see **Algorithm 2**.

In **Algorithm 2**, $P(k)$ is the $k$ th element in the input (partial) passing order, $Z_i$ is a set which consists of all cells that vehicle $i$ will pass through, $t_{\max,z}$ is the largest arrival time that cell $z$ has been occupied, $t_{arrival,i,z}$ represents the desired arrival time to Cell $z$ for vehicle $i$, $t_{\min,i,z}$ is the minimum arrival time to Cell $z$ for vehicle $i$. The flow chart of the **Algorithm 2** is shown in Fig. 5.

**Algorithm 2** Passing Order to Trajectory Interpretation

**Input:** A passing order $P$
**Output:** An objective value $J$ and $t_{arrival}$

1: **while** not all of the longitudinal position distances of vehicles meet collision avoidance conditions **do**
2:     All planned vehicles travel according to its scheduled trajectory;
3:     All unplanned vehicles follow its preceding vehicle;
4: **end while**
5: **for** each $k \in [1, length(P)]$ **do**
6:     $i = P(k)$;
7:     **if** $V_i$ goes straight **then**
8:         **for** each $z \in Z_i$ **do**
9:             $t_{arrival,i,z} = \max(t_{min,i,z}, t_{max,z} + \Delta t)$;
10:         **end for**
11:         $z_{min} = 0$;
12:     **else**
13:         Find a lane change trajectory for $V_i$ that will not cause collision with all vehicles that have higher priority in the passing order;
14:         $z_{min}$ is the cell in which $V_i$ finishes changing lanes;
15:         **for** each $z \in Z_i$ **do**
16:             **if** $z > z_{min}$ **then**
17:                 $t_{arrival,i,z} = \max(t_{min,i,z}, t_{max,z} + \Delta t)$;
18:             **else**
19:                 Update $t_{arrival,i,z}$ according to the lane change trajectory;
20:             **end if**
21:         **end for**
22:     **end if**
23:     **while true do**
24:         $a_{z,min}$ is a sufficiently big number;
25:         **for** each $z \in Z_i$ & $z > z_{min}$ **do**
26:             Calculate a constant acceleration $a_z$ with which $V_i$ can arrive at Cell $z$ at time point $t_{arrival,i,z}$;
27:             **if** $a_z < a_{z,min}$ **then**
28:                 $a_{z,min} = a_z$;
29:                 $z_{min} = z$;
30:             **end if**
31:         **end for**
32:         Update the location, velocity of $V_i$ assuming that $V_i$ has arrived at Cell $z_{min}$ with the constant acceleration $a_{z,min}$;
33:         **for** each $z \in Z_i$ **do**
34:             **if** $z \leq z_{min}$ **then**
35:                 Update $t_{arrival,i,z}$ assuming that $V_i$ drives to Cell $z$ with the constant acceleration $a_{z,min}$;
36:             **else**
37:                 Update $t_{arrival,i,z}$ by subtracting the time that $V_i$ has used to reach Cell $z_{min}$;
38:             **end if**
39:         **end for**
40:         **if** $z_{min}$ is the last cell in $Z_i$ **then**
41:             break;
42:         **end if**
43:     **end while**
44:     **for** each $z \in Z_i$ **do**
45:         $t_{max,z} = t_{arrival,i,z}$;
46:     **end for**
47: **end for**
48: $J = \sum_{i=1}^{length(P)} (t_{arrival,P(i),Z_{i,last}} - t_{min,P(i),Z_{i,last}})$;

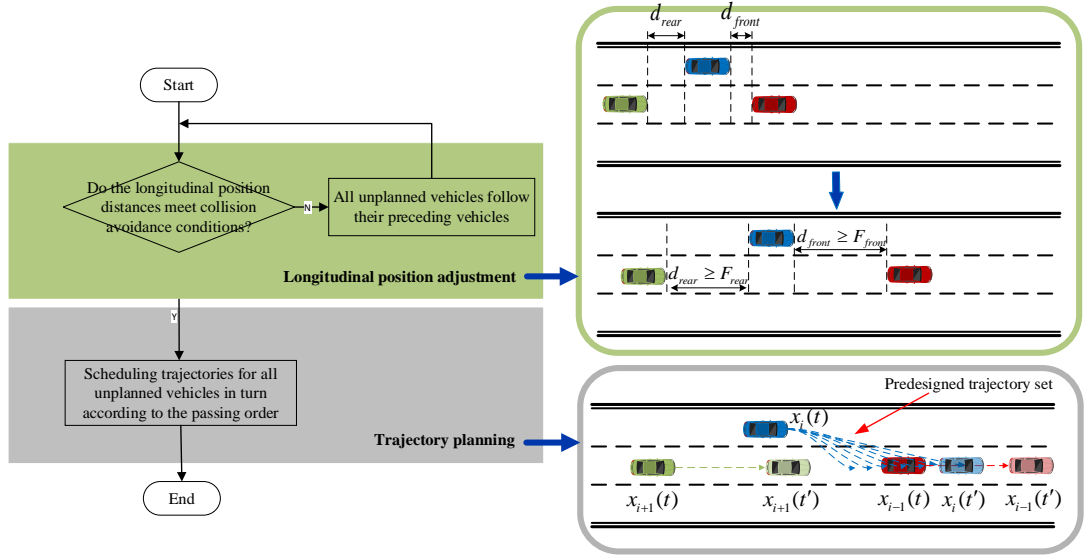

Fig. 5. The flow chart of **Algorithm 2**.

In the green part of the flow chart, we use the collision avoidance conditions proposed in (Li *et al.*, 2018) to judge whether it is safe to implement a lane change. The collision avoidance conditions are defined as:

$$d_{ij}(t) \geq F_{ij}(t)$$
$$F_{ij}(t) = v_i(t)\rho + \frac{v_i^2(t)}{2a_{i,brake}(t)} - \frac{v_j^2(t)}{2a_{j,\max,brake}} \quad (7)$$
$$a_{i,brake}(t) = a_{i,\min,brake} + \frac{v_i(t)}{v_{i,\max}}(a_{i,\max,brake} - a_{i,\min,brake})$$

where $d_{ij}(t)$ is the distance between vehicle $i$ and $j$, $F_{ij}(t)$ is the safety gap, $v_i(t)$ is the velocity of vehicle $i$, $\rho$ is the safety time headway, $a_{i,\min,brake}$ is the minimum brake acceleration, and $a_{i,\max,brake}$ is the maximum brake acceleration. We can combine various car-following models such as Newell's car-following model (Newell, 2002), the mass-spring-damper-clutch system-based car-following model (Li *et al.*, 2019) into the algorithm for guiding the unplanned vehicles to keep a safe and desired distance with its preceding vehicle.

Meanwhile, it is difficult to consider several possible lane change trajectories in the optimization problem, and all vehicles are supposed to use the same lane change trajectory. However, one candidate lane change trajectory is easy to make the problem infeasible. So, in the proposed **Algorithm 2** (line 13), some improvement is made. We first construct a trajectory set which consists of lots of different lane change trajectories that are designed according to the initial velocity and final velocity of vehicles, then let vehicles that need to change lanes search a suitable lane change trajectory from the predesigned trajectory set according to its velocity as shown in Fig. 5. Then, we update the road occupancy information according to the trajectory of the vehicle. Thus, a feasible solution of the optimization problem can be quickly obtained.

When the passing order is determined, the problem of finding corresponding trajectories for vehicles can be regarded as a degenerated case of the optimization problem (10) and it can be easily solved by **Algorithm 2**. The computation can be finished within several milliseconds. Thus, the low-level planning can quickly provide a performance evaluation of the given passing order for the upper-level planning.

## 4. Simulation Results

We design three experiments to determine the best parameter set for the new cooperative driving strategy and compare it with some classical ones. The first experiment gives a case study to explain why the bi-level-based strategy can outperform the classical cooperative driving strategy. The second experiment introduces how to determine the parameters for the proposed strategy. Finally, the third experiment compares the performance of different cooperative driving strategies.

These experiments are conducted for the work zone scenario with three lanes as shown in Fig. 1(a). The vehicles' arrival is assumed to be a Poisson process. We vary the mean value of this Poisson process to test the performance of the proposed strategy under different traffic demands.

To accurately describe the total delays of vehicles, we adopt the point-queue model in the simulation (Ban *et al.*, 2012). The model assumes vehicles travel in free flow state until they get to the boundary of the scenario we study. If the preceding vehicle leaves enough spaces, the first vehicle in the point-queue will dequeue and enter the scenario. Otherwise, it will stay in the virtual queue. Each lane has an independent point-queue.

All experiments are implemented using C++ language on a Visual Studio platform in a personal computer with an Intel i7 CPU and a 16GB RAM.

### 4.1 Case Study

To intuitively show the effectiveness of the proposed strategy, we first give a case study for a work zone scenario with 7 CAVs. The bi-level-based strategy and FIFO-based strategy are applied respectively to plan the trajectories for these vehicles, and the results are shown in Fig. 6.

The left plot in Fig. 6 shows that the movements of vehicles on lane 2 have been greatly influenced by the lane change vehicle 1 on lane 1. They need to decelerate to leave enough space for the lane change vehicle, which leads to a bigger passing time. In contrast, the bi-level-based strategy searches approximately 200 nodes within 0.5s and finds a better passing order in which vehicle 2 and vehicle 3 go first and then vehicle 1 changes its lane. Besides, the bi-level-based strategy finds it would be better to let vehicle 6 on lane 2 change its lane to lane 3 to leave more space for lane change vehicle 7. It is clear that the traffic efficiency is improved by a better passing order and the trajectories of vehicles tend to be more smooth which helps to reduce energy consumption.

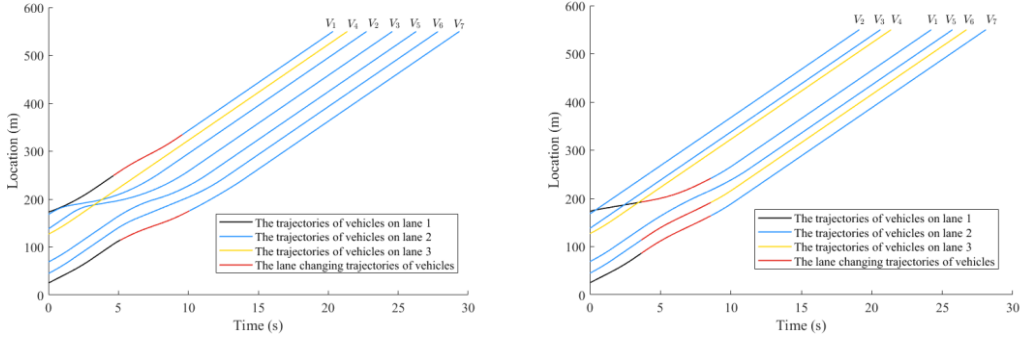

Fig. 6. (left) The planned trajectories of vehicles by the FIFO-based strategy. (right) The planned trajectories of vehicles by the bi-level-based strategy.

**4.2 The Choice of Parameters**

The parameters in the strategy undoubtedly influence its performance. For example, a bigger computation time allows us to search more nodes and thus produce better results. However, the computation time should be as small as possible in practical applications. Besides, it is important to address the problem of balancing the exploration and exploitation. In the MCTS, $\omega$ and $C$ are two weighting parameters related to the problem. A larger $C$ encourages more exploration, and $\omega$ indicates that we are more concerned about the objective values of the current node or the best simulated off-spring passing order generated from the current node. More detail introduction in terms of these parameters can be found in (Xu *et al.*, 2019b).

Since the FIFO-based passing order is the one most common used, we use it as a baseline solution and define the improvement ratio $\eta$ of the objective value when being compared with it as

$$\eta = \frac{J_{FIFO} - J_{Bi-Level}}{J_{FIFO}} \tag{21}$$

where $J_{FIFO}$ is the objective value of the FIFO-based passing order, and $J_{Bi-Level}$ is the objective value of the best passing order found by the bi-level-based strategy.

We first fix the computation time as 0.5 s and vary $\omega$ and $C$ from 0 to 1 to discuss the influence of the exploration and exploitation. The average improvement rates for the work zone scenarios with 10 vehicles are shown in Fig. 7.

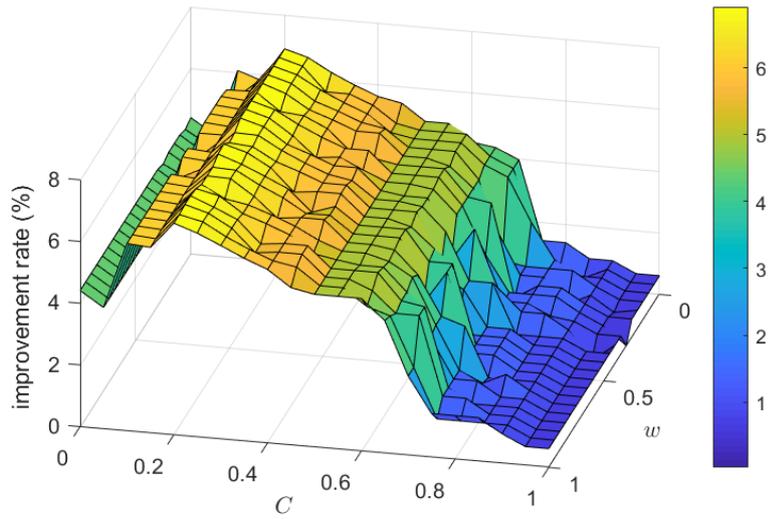

Fig. 7. The improvement rate of the bi-level-based strategy with different parameter settings.

It is clear that the parameter $C$ has an obvious impact on the improvement rates while the parameter $\omega$ is not so critical. A larger $C$ leads to more exploration and sometimes a worse result since we have wasted too much computation resource on exploring useless nodes. But it is necessary to do some exploration because the improvement rates with $C = 0.2$ are better than that with $C = 0$.

Then, we further study the scenarios with other numbers of vehicles and the results are similar. It should be noted that the improvement rates for scenarios with a small number of vehicles are small since the FIFO rule works well for these simple scenarios. Based on these results, in the rest of the paper, we set $C = 0.2$ and $\omega = 0.2$.

To further determine the appropriate maximum computation time, we investigate the relationship between the improvement rates and the number of searched nodes to eliminate the influence of computational power of experimental devices. We select the best parameter combination shown in Fig. 7 for this experiment. To better understand the performance of the strategy under different traffic demands, we vary the vehicle arrival rates to generate a series of driving scenarios with different numbers of vehicles and record corresponding improvement rates and the number of searched nodes.

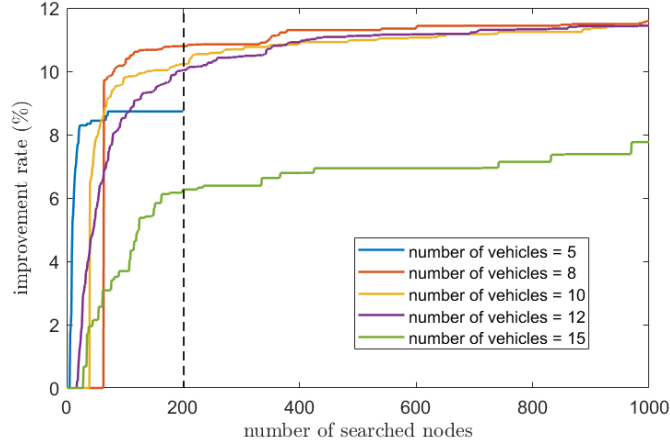

Fig. 8. The results of the improvement rates with respect to the number of searched nodes. Since the number of possible nodes is small when the number of vehicles is 5, its curve ends at abscissa 200.

Fig. 8 shows that the improvement rate increases significantly when the number of searched nodes increases from 1 to 200, and the improvement rate gets saturated afterward. Thus, it illustrates that a good enough passing order always can be found through searching 200 possible nodes for the considered scenarios. Since our experimental device can search 200 nodes within 0.5s, we set the computation time as 0.5s, and it is small enough to be applied in practice.

Besides, when the number of vehicles is small or large, the improvement rate is relatively small. This is because when the number of vehicles is small, the vehicles may be far away from each other, and FIFO-based passing order tends to be the best passing order in these simple driving scenarios; when the number of vehicles is large, there is too little space on the road to adjust the passing order.

**4.3 Comparison of Different Cooperative Driving Strategies**

In this experiment, we vary the traffic arrival rates to compare the performance of different cooperative driving strategies under different traffic demands. For each arrival rate, we simulate a 10-minute traffic process and use the bi-level-based strategy and FIFO-based strategy to coordinate the traffic process respectively. The comparison results are shown in Table 1.

To compare the coordination performance of different cooperative driving strategies, we consider two performance indices: the average delay of vehicles and the traffic throughput (the number of vehicles that has passed the work zone) within a given time interval. The delay of vehicle $i$ is defined as

$$t_{i,\text{delay}} = t_{i,\text{passing}} - t_{i,\text{min,passing}} \tag{22}$$

where $t_{i,\text{passing}}$ is the actual passing time of vehicle $i$ and $t_{i,\text{min,passing}}$ is the minimum passing time of vehicle $i$.

Table 1 Comparison results of different cooperative driving strategies

| Arrival rate (veh/h) | Strategies | Throughput (veh) | Average delay (s) |
|---|---|---|---|
| 360 | Bi-Level | 51 | 0.1528 |
| 360 | FIFO | 51 | 0.1528 |
| 1200 | Bi-Level | 205 | 0.5981 |
| 1200 | FIFO | 205 | 0.7972 |
| 1800 | Bi-Level | 296 | 0.9695 |
| 1800 | FIFO | 295 | 1.7412 |
| 2400 | Bi-Level | 382 | 3.3484 |
| 2400 | FIFO | 372 | 6.8925 |

The results in Table 1 show that the throughputs under low arrival rates are similar, but benefits in throughput can be obtained under high arrival rates. This is because there are a few vehicles in the control zone under low traffic demands, and vehicles still can pass the work zone area even if the cooperative strategy is inefficient. However, the inefficient cooperative strategy results in a bigger average delay. The differences between the average delay of two cooperative strategies increase with the arrival rates, and the delay can be effectively reduced when the arrival rate is high. Thus, the bi-level-based cooperative strategy is a promising way of traffic coordination in the future.

5. **Conclusion**

In this paper, a novel cooperative driving strategy is proposed to improve the safety and traffic efficiency for driving scenarios allowing lane changes. According to the conflict zone classification, we design a bi-level-based strategy in which the right-of-way for vehicles is considered in the upper-level, and the right-of-way of vehicles during the lane changes are solved in the lower-level. For the upper-level planning, we construct a tree representation for possible solutions and use the MCTS with some heuristic rules to accelerate the search process. For the lower-level planning, we design a passing-order-to-trajectory interpretation algorithm to quickly derive trajectories for vehicles and feedback the corresponding objective value of the given passing order to the upper-level planning. Finally, the effectiveness of the proposed strategy is validated through simulation experiments which show that the traffic efficiency is improved when compared with the classical cooperative driving strategy.

It should be mentioned that the proposed strategy can be easily extended to other driving scenarios with arbitrary road geometry. However, due to the space limit, the influence of the penetration rate of CAVs on the strategy has not been covered in the paper. We will leave this topic for our future studies. Besides, we are currently building several CAVs prototypes so that we can test our strategy in field studies in the future.

*Acknowledgements*

This work was supported in part by the National Key Research and Development Program of China under Grant 2018YFB1600600, National Natural Science Foundation of China under Grant 61673233, 61790565, 61603005, and Beijing Municipal Commission of Transport Program under Grant ZC179074Z.